\documentclass[conference]{IEEEtran}
\usepackage{flushend}

\usepackage{microtype}
\usepackage{type1cm}
\usepackage{url}
\usepackage{amsmath}
\usepackage{amsthm}
\usepackage{amssymb}
\usepackage{bm}
\usepackage{xspace}

\newtheorem{lemma}{Lemma} \newtheorem{theorem}{Theorem}
\newtheorem{corollary}{Corollary}

\theoremstyle{definition}

\newtheorem{definition}{Definition}

\theoremstyle{remark}

\newtheorem{note}{Note}

\theoremstyle{remark} \newtheorem{exampleAux}{Example}

\newenvironment{example}[1] {
  \begin{exampleAux}\label{#1}
  }{ \hfill$\vartriangleleft$
  \end{exampleAux}
}

\newcommand{\safemath}[2]{\newcommand{#1}{\ensuremath{#2}\xspace}}
\safemath{\dncA}{\mathcal{A}} \safemath{\dncB}{\mathcal{B}}
\safemath{\dncC}{\mathcal{C}} \safemath{\dncD}{\mathcal{D}}
\safemath{\dncM}{\mathcal{M}}

\safemath{\mysetN}{\mathbb{N}} \safemath{\mysetC}{\mathbb{C}}
\safemath{\mysetR}{\mathbb{R}} \safemath{\mysetRP}{\mysetR^\oplus}
\safemath{\emptystring}{\varepsilon}
\safemath{\eval}{\bigl.\bigr|}
\safemath{\iof}{\mathrm{i.o.}}
\safemath{\aev}{\mathrm{a.e.}}
\newcommand{\roc}{r.o.c.\@\xspace}

\newcommand{\gf}[2]{\mathrm{G}_{#1}(#2)}
\newcommand{\cgf}[2]{\mathrm{F}_{#1}(#2)}
\newcommand{\repart}[1]{\Re\left\lbrace #1\right\rbrace}

\title{Capacity of General Discrete Noiseless Channels}

\IEEEoverridecommandlockouts

\author{\IEEEauthorblockN{G. Bocherer}
\IEEEauthorblockA{Institute for Theoretical Information
Technology\\
RWTH Aachen University\\
52056 Aachen, Germany\\ Email: boecherer@ti.rwth-aachen.de}
\and
\IEEEauthorblockN{V.C. da Rocha Jr.,
C. Pimentel}
\IEEEauthorblockA{
Communications Research Group - CODEC\\
Department of Electronics and Systems, P.O. Box 7800\\
Federal University of Pernambuco\\
50711-970 Recife PE, Brazil\\
E-mail: \{vcr,cecilio\}@ufpe.br
}
}
\begin{document}
\maketitle
\begin{abstract}
  This paper concerns the capacity of the discrete noiseless channel
  introduced by Shannon. A sufficient condition is given for the
  capacity to be well-defined. For a general discrete noiseless
  channel allowing non-integer valued symbol weights, it is shown that
  the capacity---if well-defined---can be determined from the radius
  of convergence of its generating function, from the smallest
  positive pole of its generating function, or from the rightmost real
  singularity of its complex generating function.  A generalisation is
  given for Pringsheim's Theorem and for the Exponential Growth
  Formula to generating functions of combinatorial structures with
  non-integer valued symbol weights.
\end{abstract}
\section{Introduction}
When modelling digital communication systems, there are situations
where we do not explicitly model physical noise. We rather introduce
constraints on the allowed system configurations that minimise the
influence of undesired effects. An example is the runlength-limited
constraint in magnetic recording \cite{Marcus2001}. We consider in this
paper the discrete noiseless channel (DNC) as introduced by Shannon
\cite{Shannon1948}. A DNC is specified by a set of constraints imposed
on strings over a certain alphabet, and only those strings that
fulfil the constraints are allowed for transmission or storage.  A
DNC allows the specification of two types of constraints. The first
constraint is on symbol constellations (for example, only binary
strings with not more than two consecutive $0$s are allowed), and the
second constraint is on symbol weights (for example, the symbol $a$
has to be of duration $5.53$ seconds).  Depending on the system we
want to model, the symbol weights represent the critical resource over
which we want to optimise. This can for example be duration, length or
energy. We then ask the following question. What is the maximum rate
of data per string weight that can be transmitted over a DNC?

This question was first answered by Shannon in \cite{Shannon1948}. In
\cite{Khandekar2000}, the authors extend Shannon's results to DNCs
with non-integer valued symbol weights. In both \cite{Shannon1948} and
\cite{Khandekar2000}, the authors use the following approach to derive
the capacity of a DNC. They restrict the class of considered DNCs to
those that allow the transmission of a set of strings forming a
regular language. The regularity allows then to represent the DNC by a
finite state machine and results from matrix theory are applied to
derive the capacity of the DNC.

Our approach is different in the following sense. We consider general
DNCs with the only restriction that the capacity as defined in
\cite{Shannon1948} and generalised in \cite{Khandekar2000} has to be
well-defined, which will turn out to be a restriction on the set of
possible string weights.

This allows us then to represent the combinatorial complexity of a DNC
by a generating function with a well-defined radius of convergence and
we use analytical methods to derive the capacity. In this sense, our
work is a generalisation of \cite{Khandekar2000}. Perhaps more
important, in many cases that could be treated by the techniques
proposed in \cite{Khandekar2000}, it is much simpler to construct the
generating function of the considered DNC and to use our results to
derive the capacity. We give two simple examples that may serve as
illustrations. In this sense, our work can also be considered as an
interesting alternative to \cite{Khandekar2000}.

\section{Definitions}
We formally define a DNC and its generating function as follows.
\begin{definition}
  A DNC $\dncA=(A,w)$ consists of a countable set $A$ of strings
  accepted by the channel and an associated weight function $w\colon
  A\mapsto\mysetRP$ (\mysetRP denotes the nonnegative real numbers)
  with the following property. If $s_1,s_2\in A$ and $s_1s_2\in A$
  ($s_1s_2$ denotes the concatenation of $s_1$ and $s_2$), then
  \begin{align}
    w(s_1s_2)=w(s_1)+w(s_2).
  \end{align}
  By convention, the empty string $\emptystring$ is always of weight
  zero, i.e., $w(\emptystring)=0$.
\end{definition}
\begin{definition}\label{def:generatingSeries}
  Let $\dncA=(A,w)$ represent a DNC. We define the \emph{generating
    function} of $\dncA$ by
  \begin{align}
    \gf{\dncA}{y}&=\sum\limits_{s\in A}y^{w(s)}\qquad y\in\mysetR.
  \end{align}
\end{definition}
We order and index the set of possible string weights $w(A)$ such that
$w(A)=\lbrace w_k\rbrace_{k=1}^\infty$ with $w_1<w_2<\dotsb$. We can
then write
\begin{align}
  \gf{\dncA}{y}=\sum\limits_{k=1}^\infty N[w_k]y^{w_k}
\end{align}
where for each $k\in\mysetN$, the coefficient $N[w_k]$ is equal to the
number of distinct strings of weight $w_k$. Since the coefficients
$N[w_k]$ result from an enumeration, they are all nonnegative. Note
that for any DNC \dncA, we have $\gf{\dncA}{0}=N[0]=1$ since every DNC
allows the transmission of the empty string and since there is only
one empty string.

The maximum rate of data per string weight that can be transmitted
over a DNC is given by its capacity. We define capacity in accordance
with \cite{Shannon1948} and \cite{Khandekar2000} as follows.
\begin{definition}\label{def:capacity}
  The \emph{capacity} $C$ of a DNC $\dncA=(A,w)$ is given by
  \begin{align}
    C=\underset{k\rightarrow\infty}{\lim\sup}\dfrac{\ln N[w_k]}{w_k}
  \end{align}
  in nats per symbol weight. This is equivalent to the following. For
  all $\epsilon$ with $C>\epsilon>0$, the following two properties
  hold.
  \begin{enumerate}
  \item The number $N[w_k]$ is greater than or equal to $e^{w_k(C-\epsilon)}$
    infinitely often (\iof) with respect to $k$.
  \item The number $N[w_k]$ is less than or equal to $e^{w_k(C+\epsilon)}$
    almost everywhere (\aev) with respect to $k$.
  \end{enumerate}
\end{definition}
We assume in the following that the number sequence $\lbrace
w_k\rbrace_{k=1}^\infty$ is not too dense in the sense that for any
integer $n\geq 0$
\begin{align}
  \underset{w_k<n}{\max}k\leq Ln^K\label{eq:notTooDense}
\end{align}
for some constant $L\geq 0$ and some constant $K\geq 0$. Otherwise,
the number of possible string weights in the interval $[n,n+1]$
increases exponentially with $n$. In this case, Definition
\ref{def:capacity} does not apply. We present in the following example
a case where capacity is not well-defined.
\begin{example}{ex:tooDense}
  Let $N[w_k]$ denote the coefficients of the generating function of
  some DNC. Assume $N[w_k]=1$ for all $k\in\mysetN$ and assume
  \begin{align}
    \underset{w_k<n}{\max}k=\left\lceil R^n\right\rceil
  \end{align}
  for some $R>1$. According to Definition \ref{def:capacity}, the
  capacity of the DNC is then equal to zero because of $\ln N[w_k]=0$ for
  all $k\in\mysetN$. However, the channel accepts $R^n$ distinct
  strings of weight smaller than $n$. The average amount of data per
  string weight that we can transmit over the channel is thus
  lower-bounded by $\ln R^n/n=\ln R$, which is according to the assumption
  greater than zero.
\end{example}
Whenever we say that the capacity of a DNC is well-defined, we mean
that \eqref{eq:notTooDense} is fulfilled.
\section{Capacity by Radius of Convergence}
One way to calculate the capacity of a DNC is by determining the
radius of convergence of its generating function.
\begin{lemma}\label{lem:capacityConvergence}
  Let $\dncA$ be a DNC with the generating function $\gf{\dncA}{y}$.
  If the capacity $C$ of \dncA is well-defined, then it is given by
  $C=-\ln R$ where $R$ denotes the radius of convergence of
  $\gf{\dncA}{y}$.
\end{lemma}
In the proof of this lemma, we will need the following result from
\cite{Khandekar2000}.
\begin{lemma}\label{lem:notTooDense}
  If \eqref{eq:notTooDense} is fulfilled and if $\rho$ is a positive
  real number, then $\sum_{k=1}^\infty \rho^{w_k}$ converges iff
  $\rho<1$.
\end{lemma}
\begin{IEEEproof}[Proof of Lemma \ref{lem:capacityConvergence}]
  We define $M[k]=N^{1/{w_k}}[w_k]$ and write $\gf{\dncA}{y}$ as
  \begin{align}
    \gf{\dncA}{y}=\sum\limits_{k=1}^\infty \bigl(M[k]y\bigr)^{w_k}.
  \end{align}
  We define the two sets $D(y)$ and $E(y)$ as
  \begin{align}
    D(y)&=\bigl\lbrace k\in\mysetN\eval
    M[k]y<1\bigr\rbrace\\
    E(y)&=\mysetN\setminus D(y)=\bigl\lbrace k\in\mysetN\eval
    M[k]y\geq 1\bigr\rbrace
  \end{align}
  and write
  \begin{align}
    \gf{\dncA}{y}&=\sum\limits_{k\in D(y)}\!\!\bigl(M[k]y\bigr)^{w_k}+
    \sum\limits_{l\in E(y)}\!\!\bigl(M[l]y\bigr)^{w_l}.
  \end{align}
  It follows from Lemma \ref{lem:notTooDense} that $\gf{\dncA}{y}$
  converges iff the set $E(y)$ is finite. The number $R$ is the radius
  of convergence of $\gf{\dncA}{y}$, therefore, for any $\delta>1$,
  the set $E(R/\delta)$ is finite. Since $D(y)=\mysetN\setminus E(y)$,
  the finiteness of $E(R/\delta)$ is equivalent to
  \begin{align}
    k\in D(R/\delta)\qquad\aev\label{eq:lemma1:1}
  \end{align}
  We define $\epsilon=\ln \delta$. Equation \eqref{eq:lemma1:1} is
  then equivalent to
  \begin{align}
    N[w_k]<e^{w_k(-\ln R+\epsilon)}\qquad\aev
  \end{align}
  which implies
  \begin{align}
    N[w_k]\leq e^{w_k(-\ln R+\epsilon)}\qquad\aev\label{eq:lemma1:2}
  \end{align}
  Again since $R$ is the radius of convergence of $\gf{\dncA}{y}$, for
  any $\delta>1$, the set $E(R\delta)$ is infinite. For $\epsilon =
  \ln\delta$, this is equivalent to
  \begin{align}
    N[w_k]\geq e^{w_k(-\ln R-\epsilon)}\qquad\iof\label{eq:lemma1:3}
  \end{align}
  It follows from \eqref{eq:lemma1:2} and \eqref{eq:lemma1:3} and
  Definition \ref{def:capacity} that $-\ln R$ is equal to the capacity of
  $\dncA$. We therefore have $C=-\ln R$.
\end{IEEEproof}
In the following example, we show how Lemma
\ref{lem:capacityConvergence} applies in practice. We denote by $A\cup
B$ the union of the two sets $A$ and $B$, we denote by $AB$ the set of
all concatenations $ab$ with $a\in A$ and $b\in B$, and we denote by
$S^\star$ the Kleene star operation on $S$, which is defined as
$S^\star=\epsilon\cup S\cup SS\cup\dotsb$.
\begin{example}{ex:capacityConvergence}
  We consider a DNC $\dncA=(A,w)$ with the alphabet $\lbrace
  0,1\rbrace$ and symbol weights $w(0)=1$ and $w(1)=\pi$. The DNC
  \dncA does not allow strings that contain two or more consecutive
  $1$s. We represent $A$ by a regular expression and write
  $A=\lbrace\emptystring\cup 1\rbrace\lbrace 0\cup 01\rbrace^\star$.
  For the generating function of \dncA we get
  \begin{align}
    \gf{\dncA}{y}&=(1+y^\pi)\sum\limits_{n=0}^\infty (y+y^{1+\pi})^n.
  \end{align}
  The radius of convergence is given by the smallest positive solution
  $R$ of the equation $y+y^{1+\pi}=1$. We find $R=0.72937$. According
  to Lemma \ref{lem:capacityConvergence}, the capacity of \dncA is
  thus given by $C=-\ln R=0.31558$.
\end{example}
\section{Capacity by Rightmost Real Singularity}
There are cases where we derive the closed-form representation of the
generating function of a DNC without explicitly using its series
representation. The techniques introduced in \cite{Pimentel2003} and
\cite{Guibas1981} may serve as two examples. In this section, we show
how the capacity of a DNC \dncA can be determined from the closed-form
representation of its generating function. We do this in two steps. We
first identify the region of convergence (\roc) of the complex
generating function $\cgf{\dncA}{e^{-s}}$ with its rightmost real
singularity. The complex generating function $\cgf{\dncA}{e^{-s}}$
results from evaluating the generating function $\gf{\dncA}{y}$ in
$y=e^{-s}$, $s\in\mysetC$. Second, we show that the rightmost real
singularity of $\cgf{\dncA}{e^{-s}}$ determines the capacity of \dncA.
\begin{theorem}\label{theo:pringsheim}
  If the \roc of $\cgf{\dncA}{e^{-s}}$ is determined by $\repart{s}>Q$,
  then $\cgf{\dncA}{e^{-s}}$ has a singularity in $s=Q$.
\end{theorem}
\begin{IEEEproof}
  Suppose in contrary that $\cgf{\dncA}{e^{-s}}$ is analytic in $s=Q$
  implying that it is analytic in a disc of radius $r$ centred at
  $Q$. We choose a number $h$ such that $0<h<r/3$, and we consider the
  Taylor expansion of $\cgf{\dncA}{e^{-s}}$ around $s_0=Q+h$ as follows.
  \begin{align}
    &\cgf{\dncA}{e^{-s}}=\sum\limits_{n=0}^\infty
\dfrac{\bigl[\cgf{\dncA}{e^{-s_0}}\bigr]^{(n)}}{n!} (s-s_0)^n\\
    &=\sum\limits_{n=0}^\infty \dfrac{\sum\limits_{k=1}^\infty
      N[w_k](-w_k)^n e^{-w_k s_0}}{n!}(s-s_0)^n.
  \end{align}
  For $s=Q-h$, this is according to our supposition a converging
  double sum with positive terms and we can reorganise it in any way
  we want. We thus have convergence in
  \begin{align}
    \cgf{\dncA}{e^{-Q+h}}&=\sum\limits_{n=0}^\infty
    \dfrac{\sum\limits_{k=1}^\infty N[w_k](-w_k)^n e^{-w_k
        s_0}}{n!}(-2h)^n\\
    &=\sum\limits_{k=1}^\infty N[w_k]e^{-w_k
      s_0}\sum\limits_{n=0}^\infty \dfrac{w_k^n (2h)^n}{n!}\\
    &=\sum\limits_{k=1}^\infty N[w_k]e^{-w_k
      s_0}e^{w_k 2h}\\
    &=\sum\limits_{k=1}^\infty N[w_k]e^{-w_k(Q-h)}.
  \end{align}
  But convergence in the last line contradicts that the \roc of
  $\cgf{\dncA}{e^{-s}}$ is strictly given by $\repart{s}>Q$.
\end{IEEEproof}
We now relate the rightmost real singularity of $\cgf{\dncA}{e^{-s}}$ to
the capacity of \dncA.
\begin{theorem}\label{theo:capacity}
  Assume that $\cgf{\dncA}{e^{-s}}$ has its rightmost real singularity in
  $s=Q$.  The capacity of \dncA is then given by $C=Q$.
\end{theorem}
\begin{IEEEproof}
  Since $\cgf{\dncA}{e^{-s}}$ has its rightmost real singularity in $s=Q$,
  it follows from Theorem \ref{theo:pringsheim} that the \roc of
  $\cgf{\dncA}{e^{-s}}$ is determined by $\repart{s}>Q$. For
  $\cgf{\dncA}{e^{-s}}$, we have
  \begin{align}
    \cgf{\dncA}{e^{-s}}&=\sum\limits_{k=1}^\infty N[w_k]e^{-w_k s}\\
    &\leq \sum\limits_{k=1}^\infty |N[w_k]e^{-w_k s}|\label{eq:theo2:1}\\
    &= \sum\limits_{k=1}^\infty N[w_k]|e^{-w_ks}|\label{eq:theo2:2}
  \end{align}
  where equality in \eqref{eq:theo2:2} holds because the coefficients
  $N[w_k]$ are all nonnegative and where we have equality in
  \eqref{eq:theo2:1} if $s$ is real. It follows that if the \roc of
  $\cgf{\dncA}{e^{-s}}$ is given by $\repart{s}>Q$, then the radius of
  convergence of $\gf{\dncA}{y}$ is given by $R=e^{-Q}$. Using Lemma
  \ref{lem:capacityConvergence}, we have for the capacity $C=-\ln
  R=Q$.
\end{IEEEproof}
\begin{note}
  With Theorem \ref{theo:pringsheim} and Theorem \ref{theo:capacity},
  we generalised Pringsheim's Theorem and the Exponential Growth
  Formula, see \cite{Flajolet2008}, to generating functions of DNCs with
  non-integer valued symbol weights.
\end{note}
\section{Capacity by Smallest Positive Pole}
We formulate the most important application of Theorem
\ref{theo:capacity} in the following corollary:
\begin{corollary}\label{cor:capacity}
  Let $\dncA$ represent a DNC with a well-defined capacity $C$.
  Suppose that the generating function $\gf{\dncA}{y}$ can be written
  as
  \begin{align}
    \gf{\dncA}{y}=\dfrac{n_1y^{\tau_2}+n_2y^{\tau_2}+\dotsb+n_py{^\tau_p}}{d_1y^{\nu_1}+d_2y^{\nu_2}+\dotsb+d_qy^{\nu_q}}\label{eq:quasiPolynomial}
  \end{align}
  for some finite positive integers $p$ and $q$.  The capacity $C$ is then given by
  $-\ln P$ where $P$ is the smallest positive pole of $\gf{\dncA}{y}$.
\end{corollary}
\begin{note}
  The corollary was already stated in \cite[Theorem 1]{Pimentel2003}.
  However, the proof given by the authors does not apply for the
  general case, which we consider in this paper.
\end{note}
\begin{IEEEproof}[Proof of Corollary \ref{cor:capacity}]
  If $\gf{\dncA}{y}$ is of the form \eqref{eq:quasiPolynomial}, the
  complex generating function $\cgf{\dncA}{e^{-s}}$ as defined in the
  previous section is meromorphic, which implies that all its
  singularities are poles. The substitution $y=e^{-s}$, for $s$ real, is
  a one-to-one mapping from the real axis to the positive real axis.
  Therefore, if $Q$ is the rightmost real singularity of
  $\cgf{\dncA}{e^{-s}}$, then $e^{-Q}$ is the smallest positive pole of
  $\gf{\dncA}{y}$. Applying Theorem \ref{theo:capacity}, we get for
  the capacity $C=Q=-\ln P$.
\end{IEEEproof}
\begin{example}{ex:capacityPole}
  We consider the DNC $\dncA=(A,w)$ where $A$ is the set of all binary
  strings that do not contain the substring $111$ and where the symbol
  weights are given by $w(0)=w(1)$. We use a result from
  \cite{Guibas1981} in the form of \cite[Proposition 1.4]{Flajolet2008}.
  It states that the set of binary strings that do not contain a certain
  pattern $p$ has the generating function
  \begin{align}
    f(y)&=\dfrac{c(y)}{y^k+(1-2y)c(y)}
  \end{align}
  where $k$ is the length (in bits) of $p$ and where $c(y)$ is the
  autocorrelation polynomial of $p$. It is defined as
  $c(y)=\sum_{i=0}^{k-1}c_i y^i$ with $c_i$ given by
  \begin{align}
    c_i=\delta[p_{1+i}p_{2+i}\dotsm p_{k},p_1p_2\dotsm p_{k-i}]
  \end{align}
  where $p_i$ denotes the $i$th bit (from the left) of $p$ and where
  $\delta[a,b]=1$ if $a=b$ and $\delta[a,b]=0$ if $a\neq b$.  For
  $p=111$, we have $c(y)=1+y+y^2$ and $k=3$.  This yields for the
  generating function of \dncA
  \begin{align}
    \gf{\dncA}{y}&=\dfrac{1+y+y^2}{y^3+(1-2y)(1+y+y^2)}.
  \end{align}
  Note that the application of the technique from \cite{Pimentel2003}
  would have led to the same formula. For the smallest positive pole
  $P$ of $\gf{\dncA}{y}$ we find $P=0.54369$. According to Corollary
  \ref{cor:capacity}, the capacity of \dncA is thus given by $C=-\ln
  P=0.60938$.
\end{example}
\section{Conclusions}
For a general DNC, we identified the capacity with the characteristics
of its generating function, namely the radius of convergence of its
generating function, the rightmost real singularity of its complex
generating function, and the smallest positive pole of its generating
function. We generalised Pringsheim's Theorem and the Exponential
Growth Formula as given in \cite{Flajolet2008} to generating functions
that allow non-integer valued symbol weights.

Representing a DNC by its generating function and not by a finite
state machine has an additional advantage. Although the finite state
machine allows the derivation of the correct capacity of the DNC, it
says nothing about the exact number of valid strings of weight $w$.
The generating function of a DNC provides this information. The
coefficients $N[w_k]$ are equal to the number of distinct strings of
length $w_k$ that are accepted by the DNC. The coefficients can
either be calculated by an algebraic expansion of the generating
function or they can be approximated by means of analytic
asymptotics as discussed for integer valued symbol weights in
\cite{Flajolet2008}. In \cite{Bocherer2007a}, the analytic approach is
extended to generating functions of DNCs with non-integer valued
symbol weights.

For a regular DNC fulfilling some further restrictions, the authors in
\cite{Khandekar2000} define a Markov process that generates valid
strings at an entropy rate equal to the capacity of the channel. Based
on generating functions as introduced in this paper, it is shown in
\cite{Bocherer2007a} that for a general DNC, any entropy rate $C'$
smaller than the capacity $C$ is achievable in the sense that there
exists a random process that generates strings that are transmitted
over the channel at an entropy rate $C'$.

\section*{Acknowledgement}
V. C. da Rocha Jr. and C. Pimentel acknowledge partial support of this research by the
Brazilian National Council for Scientific and Technological
Development (CNPq) under Grants No. 305226/2003-7 and 301253/2004-8,
respectively.

\bibliographystyle{IEEEtran}

\bibliography{IEEEabrv,confs-jrnls,Literatur}

\end{document}